\documentclass[aps,prl,twocolumn,superscriptaddress,footnotebib]{revtex4-1}
\usepackage{amsmath}
\usepackage{amssymb}
\usepackage{amsfonts}
\usepackage{bm}
\usepackage{color}
\usepackage{graphicx}
\usepackage{epstopdf}
\usepackage{epsfig}
\usepackage[naturalnames]{hyperref}
\usepackage{soul}
\usepackage{hypcap}
\usepackage{verbatim}
\usepackage{tabularx}
\usepackage{bbm}
\usepackage{esvect}
\usepackage{subfigure}

\def\bea{\begin{eqnarray}}
\def\eea{\end{eqnarray}}
\def\nn{\nonumber}
\def\ba{\begin{array}}
\def\ea{\end{array}}
\def\Tr{\text{Tr}}
\def\nn{\nonumber}
\def\sgn{\text{sgn}}

\hypersetup{colorlinks=true, citecolor=blue, urlcolor=blue, linkcolor=blue}
\begin{document}
	
\title{Measurement-induced phase transition in the monitored Sachdev-Ye-Kitaev model}
	
	\author{Shao-Kai Jian}
	\thanks{They contribute equally to this work.}
	\affiliation{Condensed Matter Theory Center and Joint Quantum Institute,
Department of Physics, University of Maryland, College Park, MD 20742, USA}

	\author{Chunxiao Liu}
	\thanks{They contribute equally to this work.}
	\affiliation{Department of Physics, University of California Santa Barbara, Santa Barbara, CA 93106, USA}
	
	\author{Xiao Chen}
	\thanks{chenaad@bc.edu}
\affiliation{Department of Physics, Boston College, Chestnut Hill, MA 02467, USA}
    
    \author{Brian Swingle}
    \thanks{bswingle@umd.edu}
    \affiliation{Department of Physics, Brandeis University, Waltham, Massachusetts 02453, USA}
    \affiliation{Condensed Matter Theory Center and Joint Quantum Institute,
Department of Physics, University of Maryland, College Park, MD 20742, USA}
    
    \author{Pengfei Zhang}
	\thanks{pzhang93@caltech.edu}
	\affiliation{Institute for Quantum Information and Matter and Walter Burke Institute for Theoretical Physics, California Institute of Technology, Pasadena, CA 91125, USA}


\begin{abstract}
We construct Brownian Sachdev-Ye-Kitaev (SYK) chains subjected to continuous monitoring and explore possible entanglement phase transitions therein. We analytically derive the effective action in the large-$N$ limit and show that an 
entanglement transition is caused by the symmetry breaking in the enlarged replica space. 
In the noninteracting case with SYK$_2$ chains, the model features a continuous $O(2)$ symmetry between two replicas and a 
transition corresponding to spontaneous breaking of that symmetry upon varying the measurement rate. 
In the symmetry broken phase at low measurement rate, the emergent replica criticality associated with the Goldstone mode leads to a log-scaling entanglement entropy that can be attributed to the free energy of vortices. In the symmetric phase at higher measurement rate, the entanglement entropy 
obeys area-law scaling. In the interacting case, the continuous $O(2)$ symmetry is explicitly lowered to a discrete $C_4$ symmetry, giving rise to volume-law entanglement entropy in the symmetry-broken phase due to the enhanced linear free energy cost of domain walls compared to vortices. The interacting transition is described by $C_4$ symmetry breaking. We also verify the large-$N$ critical exponents
by numerically solving the Schwinger–Dyson equation.
\end{abstract}
\maketitle

{\it Introduction.---} Quantum dynamics can be non-unitary provided that the process occurs with a probability less than one, with the central example being measurement. Fathoming the effects of non-unitary evolution on many-body quantum states has emerged as a frontier in recent years, although the issues at play touch on the foundations of quantum physics. In particular, non-unitary evolution can lead to dramatic phase transitions in the entanglement structure of a many-body state, as in the recently discovered paradigm of the \textit{measurement-induced phase transition} realized in local random unitary circuits interspersed with measurements~\cite{li2018quantum, li2019measurement, skinner2019measurement, gullans2020dynamical,chan2019unitary}, in which the steady-state entanglement entropy changes from volume-law scaling to area-law scaling upon increasing the measurement rate. 

This transition has been observed in various settings including random Haar circuits, random Clifford circuits, Floquet quantum circuits, etc~\cite{zabalo2020critical, gullans2020scalable, li2020conformal, fan2020self, iaconis2020measurement, sang2020measurement, lavasani2021measurement, ippoliti2021entanglement, mazzucchi2016quantum}. It is continuous and enjoys an emergent conformal symmetry at the critical point. Entanglement transitions were also studied in the context of $\mathcal{PT}$ symmetry breaking, where the physics mechanism might be different~\cite{bender1998real,ashida2017parity,ashida2020non,biella2020many,gopalakrishnan2021entanglement,jian2021yang}. Moreover, this transition finds important applications in revealing phase transitions in quantum error correcting codes~\cite{choi2020quantum, gullans2020dynamical,li2020statistical}, proving efficiency/inefficiency in classical simulations of random shallow quantum circuits~\cite{napp2019efficient}, and so on. In addition to measurement-induced phase transitions in random hybrid circuits, an interesting distinct class of phenomena arise in non-interacting fermion circuits, which can host a critical phase when subjected to weak measurements. In this situation, the entanglement entropy shows a log scaling with subsystem size in an entire phase instead of just at the transition point~\cite{chen2020emergent, alberton2021entanglement, jian2020criticality, tang2021quantum, buchhold2021effective, bao2021symmetry}. As the measurement strength is further increased, a phase transition from the critical phase to the area-law phase occurs~\cite{alberton2021entanglement, bao2021symmetry}.

The phases and transitions arising from non-unitary dynamics are typically only visible in entropic observables which are non-linear in the density matrix. Accessing entropic observables averaged over the various sources of randomness thus requires averaging multiple replicas of the system with identical randomness. Depending on the observable of interest, one then has to take various kinds of replica limits. In the random Haar hybrid circuit, it is argued that the entanglement transition problem can be mapped to an order-disorder transition of a complicated statistical mechanical model in the replicated space~\cite{zhou2019emergent,skinner2019measurement, jian2020measurement, bao2020theory}. In the free fermion model, the critical phase is accounted for by a Goldstone mode resulting from the spontaneous breaking of a continuous replica rotational symmetry~\cite{buchhold2021effective, bao2021symmetry, part_one}.

\begin{figure}
	\centering
	\subfigure[]{\label{fig:chain}\includegraphics[width=0.25\textwidth]{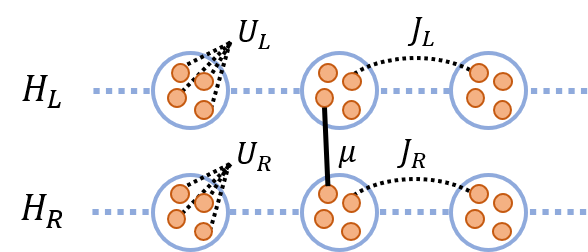}} 
	\subfigure[]{\label{fig:phase_diagram}\includegraphics[width=0.21\textwidth]{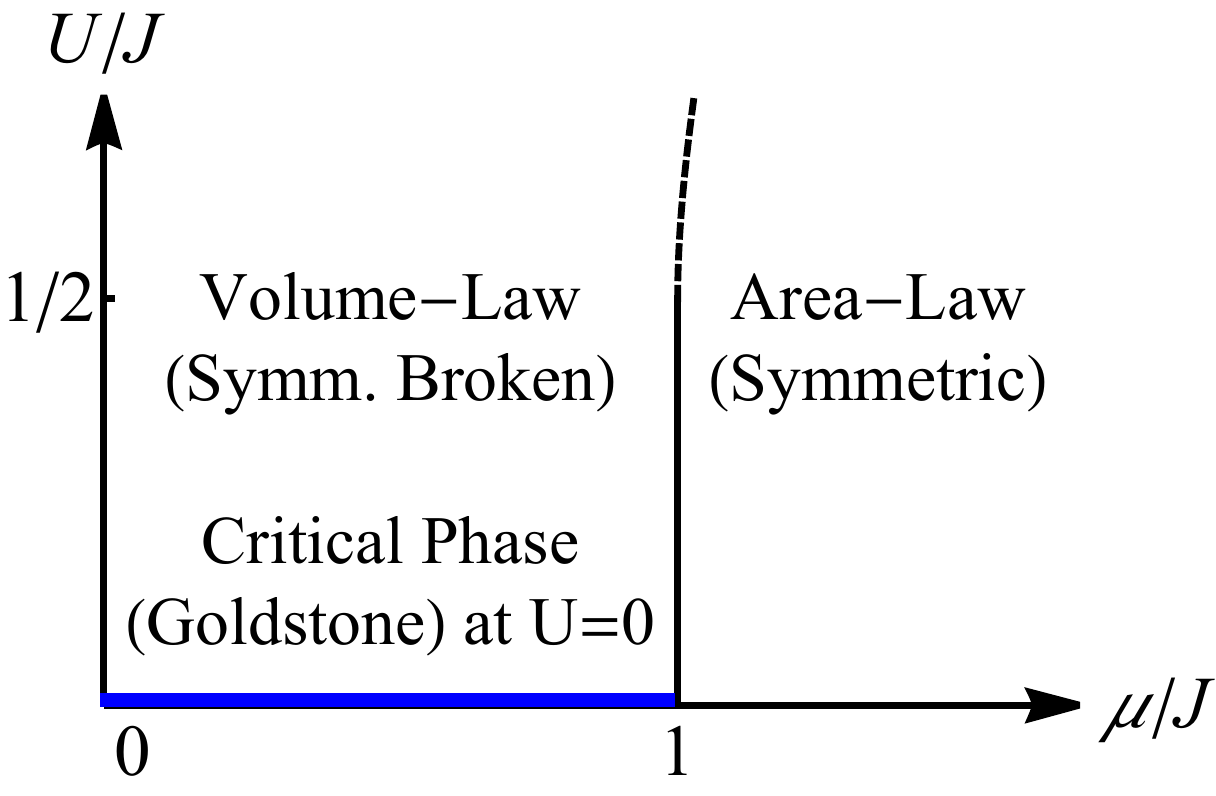}}
	\caption{(a) A schematic plot of the monitored SYK chains. The dashed (solid) line represents Brownian random couplings (monitoring operators). (b) The phase diagram of the model at infinite $N$. The black solid (dashed) line at $\mu = J$ denotes the continuous (discontinuous) transition. The blue thick line denotes the critical phase for the noninteracting case.\label{fig1}}
\end{figure}

In light of these developments, it is of great interest to construct a solvable model in which these measurement-driven transitions and critical phases can be analytically understood systematically. With this motivation in mind, we consider Brownian Sachdev-Kitaev-Ye (SYK) chains~\cite{kitaev2015simple, sachdev1993gapless, maldacena2016remarks, saad2018semiclassical, sunderhauf2019quantum, liu2021non, jian2020note} subjected to continuous monitoring, and explore possible phases and transitions. Previously, the SYK model has been extensively used to analytically understand quantum chaos and quantum information dynamics~\cite{liu2018quantum, gu2017spread, huang2019eigenstate, zhang2020subsystem, haldar2020renyi, zhang2020entanglement, chen2020replica, garcia2021replica}. We find that varying the measurement strength/monitoring rate in the non-unitary SYK dynamics causes a measurement-induced phase transition corresponding to the spontaneous breaking of $C_4$ or $O(2)$ symmetry depending on whether the model is interacting or not (see Fig.~\ref{fig1}). We extract the effective action describing the symmetry breaking and evaluate the subsystem entanglement entropy by mapping to the free energies of topological defects created by the twisted boundary conditions. We further obtain various critical exponents at the critical point and in the critical phase.

{\it Model and Setup.---} Consider the following Brownian Hamiltonian describing left ($L$) and right ($R$) chains with SYK$_2$ hoppings and SYK$_q$ on-site interactions~\cite{chen2017competition,song2017strongly},
\bea \label{eq:LRhamiltonian}
	H &= \sum_{x;a=L,R}  \Big( \sum_{ij}  i J_{a,ij}^{x,x+1}(t) \psi_{x,a,i} \psi_{x+1,a,j} \nn \\
		&+ \sum_{j_1<...<j_q} i^{q/2} U_{a,j_1...j_q}^x(t) \psi_{x,a,j_1}... \psi_{x,a,j_q}  \Big),
\eea
where $\psi_{x,a,i}$ $i=1,...,N$ denotes $i$-th of $N$ Majorana fermion at each site $x=1,...,L$ of the $a = L, R$ chains. $L$ is the number of sites and periodic boundary conditions are assumed in this paper. In the second line, $q\ge 4$ is an even integer indicating $q$-body interaction~\cite{maldacena2016remarks}. $J_{a,ij}^{x,x+1}$ ($U^x_{a,j_1,...,j_q}$) is the hopping (interaction) strength. The couplings in the left and right chains are {\it independent} Gaussian variables with mean zero and variances
\bea\label{eq:random}
& \overline{J_{a,ij}^{x,x+1}(t_1) J_{a',ij}^{x',x'+1}(t_2)} = \frac{J_a}{2N} \delta(t_{12})  \delta_{aa'} \delta^{x,x'}, \\
& \overline{U_{a,j_1...j_q}^{x}(t_1) U_{a',j_1...j_q}^{x'}(t_2)} = \frac{2^{q-2}(q-1)! U_a}{N^{q-1}} \delta(t_{12})  \delta_{aa'} \delta^{x,x'}. \nn
\eea
The time-dependence and Dirac $\delta$ functions indicate the Brownian nature of the couplings. For simplicity, we set $J_L = J_R = J $ and $U_L = U_R = U $ throughout the paper.

In addition, the system is under continuous monitoring: in each infinitesimal time step $\delta t$, we apply a measurement with probability $p$ at every site. The local measurement operator couples the $L$ annd $R$ fermions at each site, as described by the operators
\bea \label{eq:kraus}
\{ M_1^{x,i}, M^{x,i}_2\}= \left\{   \pi^-_{x,i} + \sqrt{1-s^2} \pi^+_{x,i}  ,s \pi^+_{x,i} \right\},
\eea
where $\pi^\pm_{x,i} = \frac{1}{2} (1 \mp i 2\psi_{x,L,i} \psi_{x,R,i})$ is the projection to one of the Fermi parity eigenstates and $0<s\le1$ is the measurement strength. Notice that $M^{x,i}_1$ and $M^{x,i}_2$ satisfy the required completeness relation $M_1^{x,i\dag} M^{x,i}_1+M_2^{x,i\dag} M^{x,i}_{2}=I$.

It is convenient to further introduce the Kraus operator, $K_{\mu}^{x,i} \! = \! \{ I, M_1^{x,i}, M_2^{x,i} \}$ with weights $ w_0 \! = \! (1 \!- \!p)$ , $w_1 \! = \! w_2 \! = \!p$~\cite{jian2020criticality}.
During each time step in $\delta t$, the evolution of the non-normalized density matrix for a quantum trajectory with measurement outcome $\mu_{x,i}$ is~\cite{wiseman1993quantum, carmichael1993quantum, wiseman1996quantum, gardiner2004quantum}
\bea
    && \rho_{ab}(\delta t; \mu_{x,i}) = \nn \\
    && (\otimes_{x,i} K_{\mu_{x,i}}^{x,i})_{ac} (\otimes_{x,i} K_{\mu_{x,i}}^{x,i\dag})_{db}  (e^{-i H \delta t})_{ce} (e^{i H \delta t})_{fd} \rho_{ef}.
\eea
For observables linear in density matrix, one can perform the average in $\rho$. If measurement of a specific $i$-th Majorana at site $x$ is performed, the average change of density matrix (we suppress index $x,i$), is
$ \sum_{\nu} (w_{\nu} K_\nu \rho K_\nu^\dag ) - \rho \approx - \frac{ps^2}2 \{ \pi^+, \rho \} + p s^2 \pi^+ \rho \pi^+ $,
where we keep up to $O(s^2 )$ order. Let $s=(\mu/p) \sqrt{\delta t}$, we get the Lindblad equation $\frac{d\rho}{dt} = - \frac12 \{L^\dag L, \rho \} + L \rho L^\dag$, where $L=\mu \pi^+$ is the jump operator. For observables nonlinear in density matrix, one is not able to get a linear Lindblad equation. Therefore, we use the path integral formalism detailed below.

\begin{figure}
	\centering
	\subfigure[]{\label{fig:rho_square}
		\includegraphics[width=0.18\textwidth]{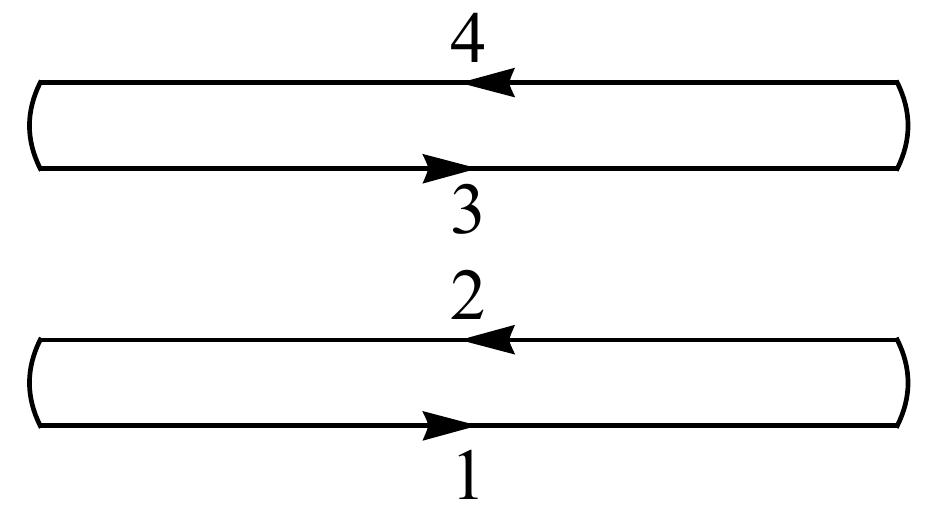}} \quad
	\subfigure[]{\label{fig:twist}
		\includegraphics[width=0.20\textwidth]{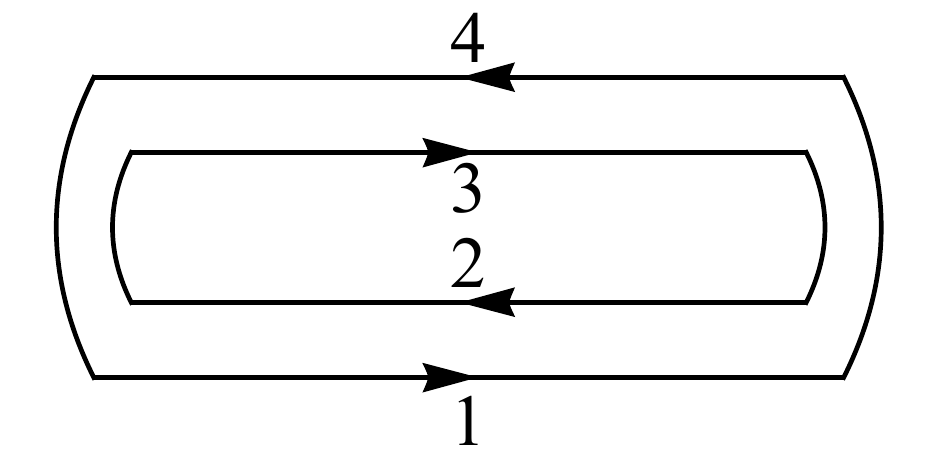}}
	\subfigure[$T\ll L_A$]{ \label{fig:2d_time}
		\includegraphics[width=0.18\textwidth]{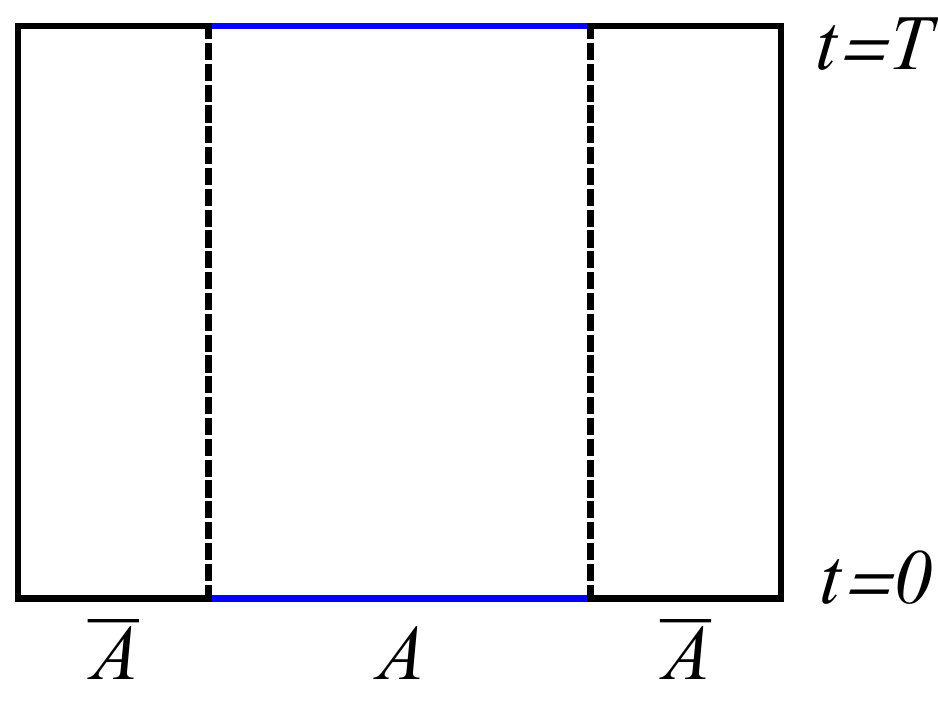}} \quad
	\subfigure[$T\gg L_A$]{ \label{fig:2d_space}
		\includegraphics[width=0.18\textwidth]{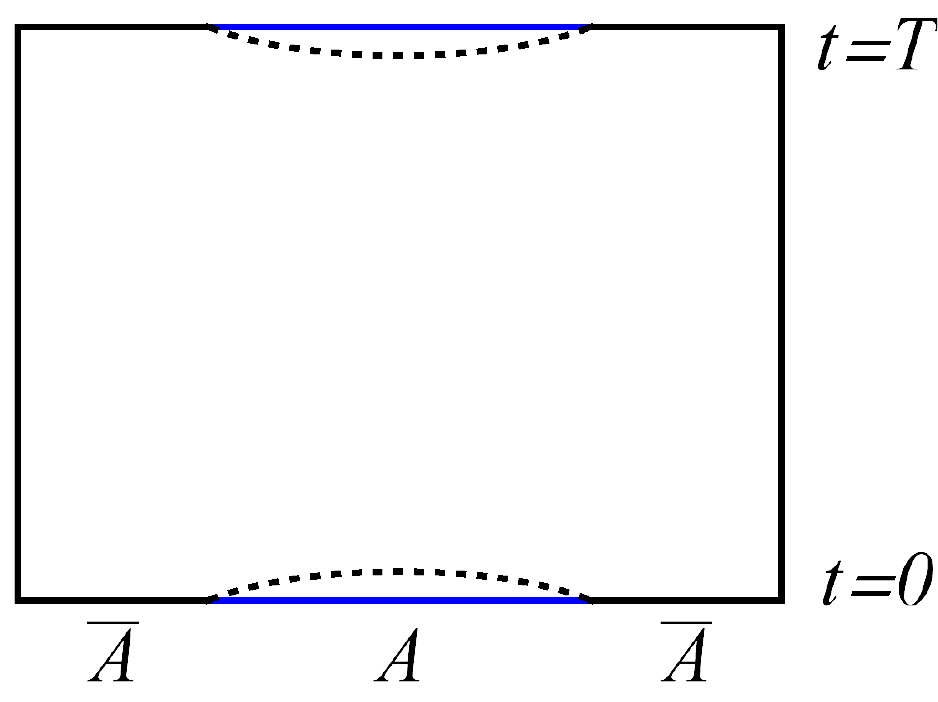}}
	\caption{ (a) The boundary condition corresponding to $\Tr(\rho)^2$. (b) The boundary condition corresponding to two twist operators inserted at $t=0$ and $t=T$, respectively. (c,d) The twisted boundary conditions are indicated by blue line in subsystem $A$. In the volume-law space, the spacetime domain wall indicated by the dashed curve separates two domains induced by different boundaries. Schematic plots of domain walls are shown at short times (c) $T \ll L_A$ and long times (d) $T \gg L_A$.}
\end{figure}

For our purpose, we are interested in calculating the quasi-$n$ entropy of bipartite system $A \bar A$~\cite{napp2019efficient},
\bea \label{eq:entropy}
	S_A^{(n)} = \frac1{1-n} \log \frac{\mathbb E \Tr(\rho_{A}^n)}{\mathbb E\Tr(\rho)^n},
\eea
where $\rho$ ($\rho_A$) is the total density matrix (the reduced density matrix of subsystem $A$ by tracing out $\bar A$), and $\mathbb E$ denotes average over the Brownian variables and the continuous monitoring overcomes. 
To evaluate this quantity, one should generalize to $n$ replicas. 
We mainly consider $n=2$, and use a $1,2,3,4$ notation: $1,2$ ($3,4$) denote the first (second) replica, and $1,3$ ($2,4$) denote the forward (backward) evolution. A schematic plot is given in Fig.~\ref{fig:rho_square} and~\ref{fig:twist}. We use superscript Greek alphabet $\alpha = 1,2,3,4$ to denote the contour.

To evaluate $\mathbb E\Tr(\rho)^2$, we derive the effective action governing the time evolution in the replicated space. 
The continuous monitoring at each step can be cast into
\bea
& \sum_\nu w_\nu (K_\nu^{x,i})^{\otimes 2} \otimes (K_\nu^{x,i\dag})^{\otimes 2}  
 \approx e^{\frac{\mu \delta t}{2}  \sum_{\alpha=1}^4 i \psi_{x,L,i}^\alpha \psi_{x,R,i}^\alpha},
\eea
which is obtained in the limit $p \ll s \ll 1$~\cite{suppl}. $\mu \equiv p s^2/\delta t$ is the relevant measurement rate that is kept fixed when the limit is taken. 
Then the effect of monitoring every Majorana species $i$ at every site $x$ is described by
\bea \label{eq:measurement}
& \exp  \left(\frac\mu2 \int dt  \sum_{x,\alpha,i} i \psi_{x,L,i}^\alpha \psi_{x,R,i}^\alpha\right),
\eea 
where we implicitly sum over all infinitesimal time steps to arrive at the time integral for a time evolution.

Combining the Brownian Hamiltonian~(\ref{eq:LRhamiltonian}) and the measurement~(\ref{eq:measurement}) and integrating out the Gaussian variables, the effective action governing the time evolution in the replicated space reads
\bea
- \frac{I}N &= \frac12 \Tr \log \left( (-1)^{\alpha+1}  \partial_t - \Sigma_{x} \right)- \frac12 \int  \Sigma_{ab,x}^{\alpha\beta} G_{ab,x}^{\alpha\beta}  \nn \\
 	& + \int \delta(t-t') \Big[ - \frac{(-1)^{\alpha + \beta}}{4} \delta_{ab} \Big( J G_{ab,x}^{\alpha\beta} G_{ab,x+1}^{\alpha\beta} \nn \\
 	 & \quad\quad\quad\quad\quad\quad + \frac{U}{2q} (2G_{ab,x}^{\alpha\beta})^{q} \Big) + \frac{i \mu}2 G_{LR,x}^{\alpha\alpha} \Big], 
\eea
where $\alpha, \beta = 1,...,4$ denote the four contours.
The summations over $x$, $a,b$ and $\alpha, \beta$ are implicit.
$\Sigma_{ab,x}^{\alpha\beta}(t,t')$ is the self energy introduced to enforce $ G_{ab,x}^{\alpha\beta}(t,t') = \frac1N \sum_j \psi_{x,a,j}^\alpha(t) \psi_{x,b,j}^\beta(t')$. Saddle-point analysis can be straightforwardly applied to the large-$N$ action~\cite{suppl}.

{\it Monitored SYK$_2$ chain and $O(2)$ transition.}--- For the noninteracting case, $U=0$, the replica diagonal spatially uniform solution (with site index suppressed) reads 
\bea \label{eq:SYK2_saddle}
\bar G = \begin{cases} \frac{e^{- \frac{J|t|}2 }}{2} \big( \sgn(t)\sigma^z - \sqrt{1-\tilde\mu^2} i \sigma^y + \tilde \mu \tau^y  \big),  \tilde \mu < 1  \\
					    \frac{e^{- \frac{\mu |t|}2 }}{2} \left( \sgn(t)\sigma^z +  \tau^y  \right), \qquad \qquad \qquad \quad \tilde \mu \ge 1 \end{cases}
\eea
where $t$ is the time difference, $\tilde \mu \equiv \mu/ J$ and Pauli matrix $\sigma $ ($\tau$) acts on 1 and 2 contours ($L$ and $R$ chains).
The solution on $3,4$ contours is the same, consistent with the boundary condition without twist operators [Fig.~\ref{fig:rho_square}]. 

First we look at the theory from symmetry perspective. For Brownian randomness~(\ref{eq:random}), it is legitimate to assume the Green functions are strictly local and antisymmetric $G^{\alpha\beta}_{ab,x}(t) \!=\! -G^{\beta\alpha}_{ba,x}(t)$~\cite{saad2018semiclassical}, so for $U\!=\!0$ the action becomes
\bea \label{eq:trace}
&& - \frac{I}N  = \frac12 \Tr \log \left( S \partial_t + \Sigma_{x} \right) \\
			&& + \int  \frac12 \Tr\Big[ \Sigma_{ab,x} G_{ba,x} + \frac{J}{4}  G_{ab,x} S G_{ba,x+1} S + i \frac{\mu}2 G_{LR,x} \Big], \nn
\eea
where $ S^{\alpha\beta} = (-1)^{\alpha} \delta^{\alpha\beta}$ and the trace in the second line is over the contours.
For a finite measurement rate $\mu\!>\!0$, the theory features $O(2) \!\times \!O(2)$ symmetry~\cite{semidirect}, 
\bea
G_{ab,x} \rightarrow O^{-1} G_{ab,x} O, \quad O^T O = 1, \quad O^T S O = S, 
\eea
where $O$ acts identically on the left and right chains~\cite{mu}. The rotational symmetry is generated by $\gamma_{(13)}$ and $\gamma_{(24)}$, which is defined in component $\alpha, \beta$ by $\gamma_{(ij)}^{\alpha\beta} =  \delta^{i\alpha} \delta^{j\beta} -\delta^{j\alpha} \delta^{i\beta}$, acting on the replica space. Intuitively, one of the $O(2)$ symmetries is to rotate between $1$ and $3$ contours and the other to rotate between $2$ and $4$ contours.

The saddle point solution~(\ref{eq:SYK2_saddle}) for $\mu<J$ spontaneously breaks the {\it relative} rotational symmetry, so there is one Goldstone mode, which is generated by applying the broken-symmetry generator $\gamma_- \equiv \gamma_{(13)} \!-\! \gamma_{(24)} $, i.e.,  
\bea \label{eq:Goldstone}
\delta G_{aa,x}(t) &= e^{-\theta_x(t) \gamma_-} \bar G_{aa}(0) e^{\theta_x(t) \gamma_-} - \bar G_{aa}(0) \\
&\approx \sqrt{1- \tilde \mu^2} \theta_x(t) (\gamma_{(14)} + \gamma_{(23)}),
\eea
where $\theta_x(t)$ denotes the Goldstone mode, and in the second line we assume the fluctuation is small, $\theta_x(t) \ll 1$. We anticipate that it will dominate at low energies at $\mu\!<\!J$. In contrast, when $\mu\!>\!J$, this $O(2)$ symmetry is unbroken and the replicated theory is in the gapped phase. 

With this understanding, we are ready to evaluate the effective action for Goldstone mode. 
First notice that $G_{LR}^{\alpha\alpha}$ is linear in the action~(\ref{eq:trace}), so it can be integrated out to enforce $\Sigma_{LR}^{\alpha\alpha} =  \frac{i \mu}2$. 
Then we consider the fluctuations $\delta \Sigma_{aa}^{\alpha\beta}$ and $\delta G_{aa}^{\alpha\beta}$ away from the saddle point solution~(\ref{eq:SYK2_saddle}) at $\mu<J$.
The effective theory for the Goldstone mode reads~\cite{suppl},
\bea
\frac{I_{\text{eff}}}{N} = \frac{\rho}2 \sum_k \int_\Omega  \left( \frac{\Omega^2}{\mu^2}  + (1-\cos k) \right) |\theta_k(\Omega)|^2,
\eea
where $\int_\Omega \!=\! \int \frac{d\Omega}{2\pi}$ and $\theta_k \!=\! \frac1{\sqrt{L}}\sum_x \theta_x e^{-i k x} $ is the Fourier transform of the lattice site. 
The stiffness $\rho \!=\! J( 1\!-\! \tilde \mu^2) $ vanishes at $\tilde \mu \! = \!1$, 
indicating that the transition occurs at $\mu \!= \!J$, which is expected because the saddle point solution restores $O(2)$ symmetry. Recently, similar mechanism was discussed in the small-$N$ case in the language of Kosterlitz–Thouless transition in Ref.~\cite{buchhold2021effective,bao2021symmetry}. The Goldstone mode also explains the power-law squared correlation function of fermions in the critical phase~\cite{suppl, part_one, chen2020emergent, alberton2021entanglement}. 

{\it Interacting SYK$_4$ model and $C_4$ transition.}---
In the interacting case $U\!>\!0$, when $\mu\!<\!J$, the uniform solution reads
\bea \label{eq:SYKq_saddle}
& \bar G = \frac{e^{- \frac{J + U \zeta^{q-2}}2 |t|}}{2} \Big[\sgn(t) \sigma^z - \zeta i \sigma^y + \frac{\tilde \mu \tau^y}{1+\tilde U \zeta^{q-2}} \Big],
\eea
with $\tilde U \equiv U/J$. The parameter $\zeta$ is given by $(1-\zeta^2)(1+ \tilde U \zeta^{q-2})^2 \!=\! \tilde \mu^2$.
$\zeta$ characterizes the correlation between forward and backward contours, and serves as an order parameter as we will see later. For small $\tilde U$, $\zeta \!=\! \sqrt{1 \!-\! \tilde \mu^2} [ 1 \!+\! \tilde \mu^2 ( 1 \!-\! \tilde \mu^2)^{q/2-2} \tilde U \!+\! O(\tilde U^2) ]$ is well defined when $\mu \!<\! J$, and vanishes continuous as $\mu \!\rightarrow\! J$. In the following we will focus on the simplest interacting case with $q=4$, while our results are true for general $q$. At the critical point, 
\bea \label{eq:x_condition}
    \zeta^2 \big((2\tilde U-1)+(\tilde U^2-2 \tilde U) \zeta^2 - \tilde U^2 \zeta^4 \big) =0,
\eea
which shows that for $\tilde U \!>\! 1/2$ there are two degenerate distinct physical solutions indicating a discontinuous jump.
Thus, the condition for a continuous transition is $ 2 U < J$. 
On the other hand, when $\mu \!\ge\! J$, the solution is the same as the noninteracting case~(\ref{eq:SYK2_saddle}) at $\tilde \mu \!\ge\! 1$.

For $U \!>\! 0$ apparently the action cannot be cast into such a nice form as~(\ref{eq:trace}), so what symmetry out of $O(2) \!\times\! O(2)$ is preserved? 
It is easy to show that the symmetry reduces to $C_4 \!\times\! C_4$, satisfying the condition $((O^{-1})^{\alpha\beta})^{q/2} S^{\beta\gamma} (O^{\gamma\delta})^{q/2} \!=\! S^{\alpha\delta} $.
The generator is still given by $\gamma_{(13)}$ and $\gamma_{(24)}$ but the rotation angle is restricted to multiples of $\pi/2$.
The relative rotation symmetry is spontaneously broken by nonzero $\zeta$ in~(\ref{eq:SYKq_saddle}) when $\mu \!<\! J$. Namely, $\zeta$ serves as an order parameter of the $C_4$ symmetry breaking transition.

\begin{figure}
	\centering
	\subfigure[]{
		\includegraphics[width=0.22\textwidth]{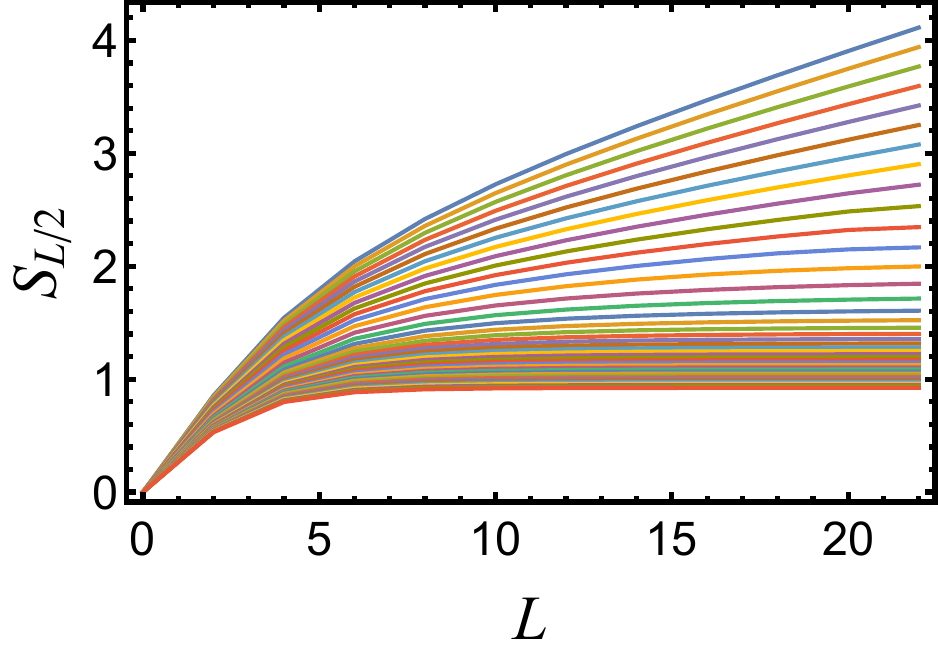}} \quad
	\subfigure[]{ \label{fig:exponent}
		\includegraphics[width=0.22\textwidth]{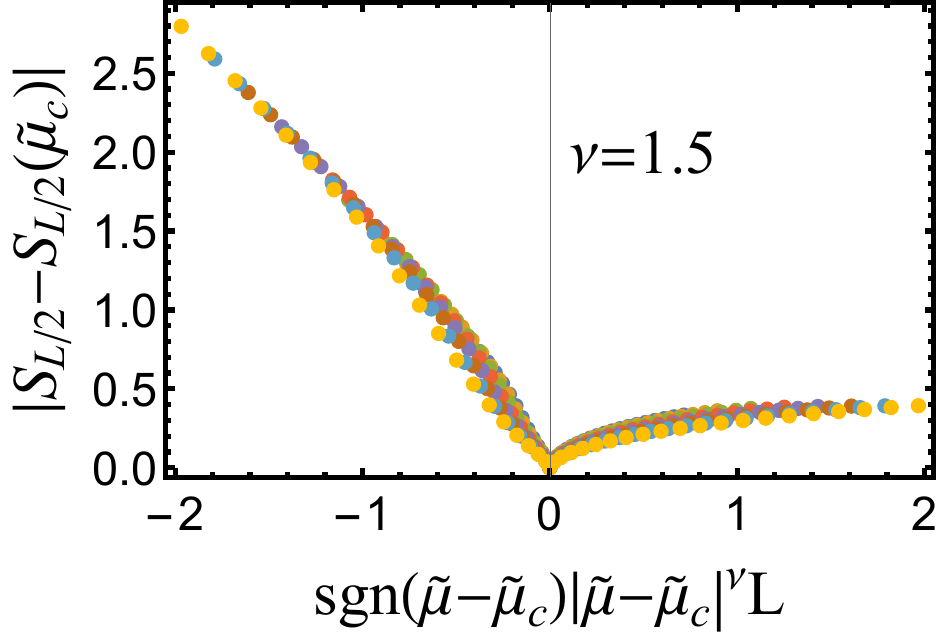}} \\
	\subfigure[]{
		\includegraphics[width=0.22\textwidth]{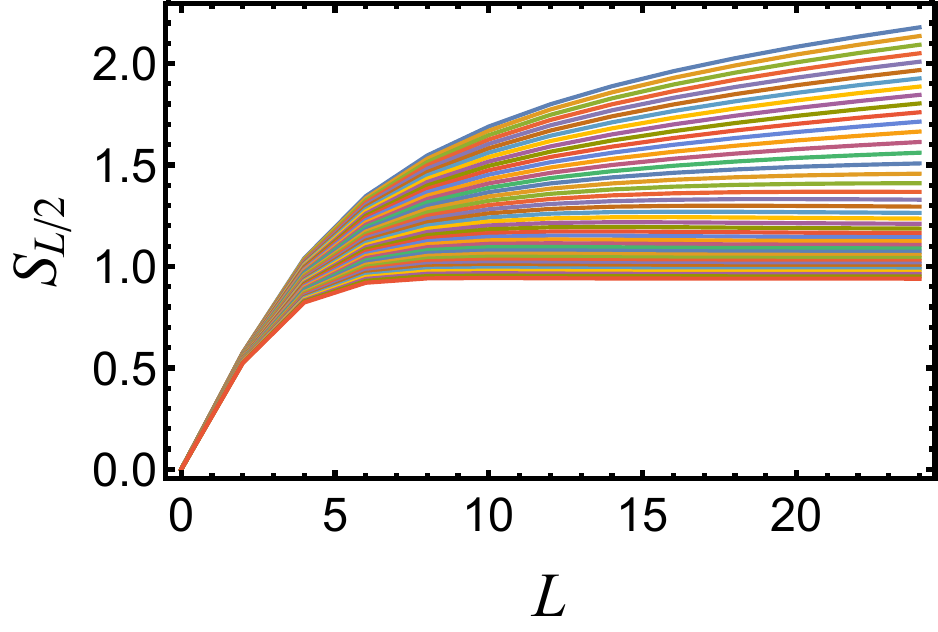}} \quad
	\subfigure[]{ \label{fig:free_exponent}
	\includegraphics[width=0.22\textwidth]{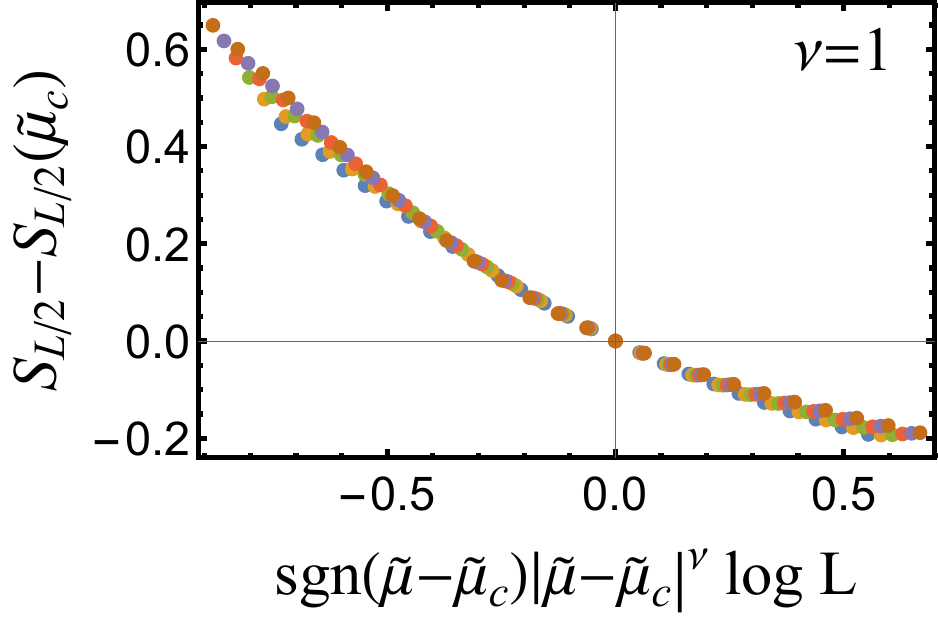}}
	\caption{The half-chain quasi entropy as a function of the length of the chain. Different curves represent different measurement strength $\tilde \mu = 0.8 ,0.81,...,1.2$ for (a) $U=0.4 J$ and (c) $U=0$. Data collapse verifies the critical exponent for (b) $U=0.4 J$ and (d) $U=0$.}
\end{figure}

Aiming at the critical theory, we consider the effective theory of fluctuations from the symmetric saddle-point solution~\cite{suppl},
\bea \label{eq:Z4_model}
\frac{I_{\text{eff}}}{N} =& \frac12 \sum_{i=1,2; k} \int_{\Omega}  \left( \frac{\Omega^2}{\mu} + J(1-\cos k) \right) |\phi_{i,k}(\Omega)|^2  \nn\\ 
 +&  \sum_x \int_{t} \left( \frac{\mu-J}2 \vec \phi_{x}^2 + \frac{\mu}8 \vec \phi_{x}^4  - \frac{U}4 ( \phi_{1,x}^4 + \phi_{2,x}^4 ) \right),
\eea
where $\int_t \!=\! \int dt$, $\phi_1 \!=\! \delta G^{12} \!+\! \delta G^{34}$ and $\phi_2 \!=\! \delta G^{14} \!+\! \delta G^{23}$~\cite{LR} transform like a vector under the relative $C_4$ rotation. 
This theory features a second order transition if $2U \!<\! \mu $, and a first order one if $2U\!>\!\mu$, consistent with the analysis~(\ref{eq:x_condition}) of the saddle-point solution. 

{\it Entanglement transition and spacetime domain wall.---} After analyzing the effective action for the averaged $\mathbb E \mbox{Tr}(\rho)^2$, we are now ready to study the entanglement transition. 
Importantly, the quasi-$n$ entropy~(\ref{eq:entropy}) tends to trajectory-averaged entanglement entropy  at $n \rightarrow 1$ limit just like the R\'enyi-$n$ entropy, which gives some justification of using quasi-2 entropy in the following as a proxy of entanglement entropy~\cite{renyi}.
To observe the entanglement transition, we start from the well-known thermofield double state (TFD) in the doubled Hilbert space~\cite{gu2017spread, penington2019replica, chen2020replica}.
The entropy of subsystem $A$ at time $T/2$ is obtained by imposing two twist operators at time $t=0$ and $t=T$ in the subsystem $A$ which change the boundary condition by requiring $G^{14}_{aa}(0) = G^{23}_{aa}(0)  =G_{aa}^{14}(T) = G^{23}_{aa}(T)  = \frac12 $ as indicated in Fig.~\ref{fig:twist}.
Equivalently in terms of the $Z_4$ model~(\ref{eq:Z4_model}) of the interacting case, the boundary of $A$ has $(\phi_1=0,\phi_2>0)$ whereas that of $\bar A$ has $(\phi_1>0,\phi_2=0)$.
In the symmetry-broken phase, it amounts to create distinct space time domains and consequently domain walls separating them as indicated in Fig.~\ref{fig:2d_time} or~\ref{fig:2d_space}. 
Then the quasi entropy is given by the free energy difference between the configurations with and without twisted boundary conditions.

\begin{figure}
    \centering
    \subfigure[]{\includegraphics[width=0.22 \textwidth]{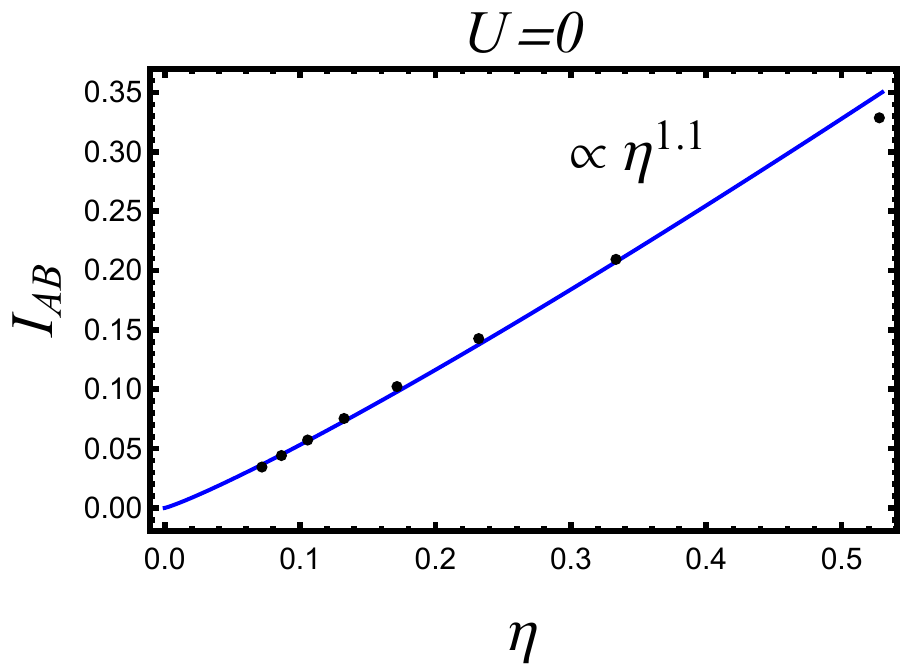}}
    \subfigure[]{\includegraphics[width=0.22 \textwidth]{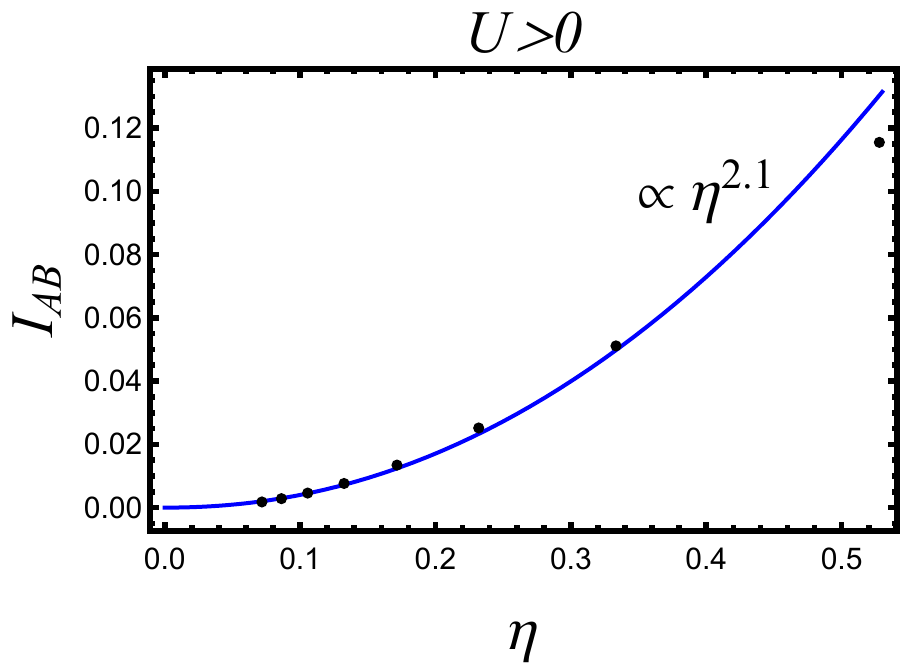}}
    \caption{Mutual information as a function of cross ratio $\eta$. The mutual information $I_{AB}$ is between $A$ and $B$, two symmetric intervals $L_A = L_B = 2$ at opposite sides in the chain of length $L=10,..,24$. (a) The free case with $\widetilde{\mu}=0.7$ in the critical phase. (b) The interacting case at the critical point. }
    \label{fig:MI}
\end{figure}

Redefining $Z_4$ the theory~(\ref{eq:Z4_model}) to be
\bea
 \frac{I_{\text{eff}}}{N} &= \int dt dx \Big( \frac12 (\partial \vec \phi)^2 + r \vec \phi^2 + \lambda \vec \phi^4 + \lambda' ( \phi_1^4 + \phi_2^4) \Big), \quad
\eea
with $ r \! = \! \frac12 \mu (\mu \!-\! J)$, $\lambda \!=\! \frac{\sqrt{2}}{8} \mu^{\frac52} J^{-\frac12}$, and $\lambda' \!=\! -\frac{2U}{\mu} \lambda $, one can find that the surface tension of domain walls reads
\bea \label{eq:tension}
\sigma = \frac{\pi J \tilde U^{\frac12}}{2\sqrt2}   \left( \frac{1-\tilde \mu}{\tilde \mu - 2 \tilde U}  \right)^{3/2}.
\eea
This can also be estimated as $\sigma \!\sim\! \xi(\phi_0^2/\xi^2) \!\sim\! r^{3/2}$, where $\phi_0 \!=\! \left|\langle\vec \phi\rangle\right| \!\sim\! \sqrt{r}$ and the correlation length $\xi \!\sim\! 1/\sqrt{r}$. With this information, the quasi entropy in the volume-law phase is given by
\bea
S_A^{(2)} = \begin{cases} 2 (N \sigma T + \log T/2),  & T \ll L_A \\
	2 (N \sigma L_A + 3/2 \log L_A), & T \gg L_A 
\end{cases},
\eea
with $L_A$ the length of subsystem $A$.
The factor of $2$ is due to the doubling of the Hilbert space in defining the TFD initial state. 
The leading order term is intuitively shown by the configurations in Fig.~\ref{fig:2d_time} and~\ref{fig:2d_space}, while the subleading $1/N$ logarithmic correction is from the transverse fluctuations. Indeed this is the capillary wave theory of the volume-law phase discussed in Ref.~\cite{li2020statistical}.
Approaching the critical point $\tilde \mu \!\rightarrow \! 1$, the surface tension~(\ref{eq:tension}) vanishes continuously with an  exponent $\nu \! =\! 3/2$, and consequently the system undergoes an entanglement phase transition.
The critical exponent is verified numerically by solving the Schwinger-Dyson equation~\cite{suppl} as shown in Fig.~\ref{fig:exponent}. 
Furthermore, the purification transition can be understood in a similar way~\cite{purification}.

This geometric picture is a systematic way to understand the connection between the volume-scaling and the log-scaling entanglement entropy in the interacting and noninteracting cases respectively:
the surface tension~(\ref{eq:tension}) vanishes when the interaction strength is zero, and the domain walls created by the twisted boundary conditions in $\phi$ field change to the vortices in $\theta$ field~\cite{part_one}. 
A similar calculation shows at the leading order
$S_A^{(2)} \!\sim \! (1\! -\! \tilde\mu) N \log L_ A$, 
that is well expected from the logarithmic free energy of vortices. The critical exponent $\nu \!=\! 1$ is verified numerically in Fig.~\ref{fig:exponent}. In addition, we numerically calculate the mutual information $I_{AB}\!=\!S^{(2)}_{AB}\!-\!S^{(2)}_A\!-\!S^{(2)}_B$ between two intervals $A=[x_1,x_2]$ and $B=[x_3,x_4]$, where $x_i$ denotes the site, in the SYK chain with periodic boundary (see Fig.~\ref{fig:MI}) and we observe that it is a function of cross-ratio $\eta \!=\! \frac{\sin \frac\pi{L} x_{12} \sin \frac\pi{L} x_{34} }{\sin \frac\pi{L} x_{14} \sin \frac\pi{L} x_{23}}$ with $x_{ij} \!=\! |x_i \!-\! x_j|$. In particular, we have $I_{\text{AB}}\!\sim\! \eta^\Delta$ in the limit $\eta \!\to\! 0$. In the critical phase of the free system, $\Delta\!\approx\! 1$ and at the critical point of the interacting system, $\Delta \!\approx\! 2$. We observe that these critical exponents are consistent with the previous numerical results in the small-$N$ case~\cite{li2019measurement,chen2020emergent,skinner2019measurement}.

{\it Conclusions.---} To summarize, we investigate the measurement-induced entanglement phase transition in Brownian SYK chains in the large-$N$ limit. We show that the dynamical symmetry in the replica space plays a crucial role: a $O(2)$ symmetry underlies the physics of noninteracting cases, and it is lowered to $C_4$ by finite interactions. The entanglement entropy in both cases can be understood in a unified framework by mapping to free energy costs of topological defects created by the twisted boundary conditions.  

{\it Acknowledgement.---} We acknowledge helpful discussions with Ehud Altman, Yimu Bao, Subhayan Sahu, and Greg Bentsen. SKJ and BGS are supported by the Simons Foundation via the It From Qubit Collaboration. The work of BGS is also supported in part by the AFOSR under grant number FA9550-19-1-0360. CL is supported by the NSF CMMT program under Grants No. DMR-1818533. PZ acknowledges support from the Walter Burke Institute for Theoretical Physics at Caltech. We acknowledge the University of Maryland High Performance Computing Cluster (HPCC).

\bibliography{reference.bib}

\begin{thebibliography}{61}
\bibitem [{sem()}]{semidirect}%
  \BibitemOpen
  \href@noop {} {}\bibinfo {note} {The symmetry is $(O(2)\times O(2)) \rtimes
  Z_2$. Since the $Z_2$ remains intact in the discussion, we simply neglect
  this semidirect product structure.}\BibitemShut {Stop}%
\bibitem [{mu()}]{mu}%
  \BibitemOpen
  \href@noop {} {}\bibinfo {note} {Without the coupling between the left and
  the right chains, $\mu=0$ , the action is invariant under two $O(2) \times
  O(2)$ for the left and the right chains, respectively, $G_{ab,x} \rightarrow
  O^{-1}_a G_{ab,x} O_b$, where $O^T_a O_a = 1, O^T_a S O_a = 1$.}\BibitemShut
  {Stop}%
\bibitem [{LR()}]{LR}%
  \BibitemOpen
  \href@noop {} {}\bibinfo {note} {In this case we have $\delta G^{12}_{RR} =
  \delta G^{12}_{LL}$, $\delta G^{34}_{RR} = \delta G^{34}_{LL}$, $\delta
  G^{14}_{RR} = \delta G^{14}_{LL}$, $\delta G^{23}_{RR} = \delta
  G^{23}_{LL}$~\cite{suppl}. So we omit the subscript of the left and right
  chains.}\BibitemShut {Stop}%
\bibitem [{ren()}]{renyi}%
  \BibitemOpen
  \href@noop {} {}\bibinfo {note} {The trajectory-averaged R\'enyi entropy can
  be defined as $\tilde S^{(n)}_A = (1-n)^{-1} \mathbb E \Tr \rho \log
  \frac{\Tr \rho_A^n}{\Tr \rho^n}$}\BibitemShut {NoStop}%
\bibitem [{pur()}]{purification}%
  \BibitemOpen
  \href@noop {} {}\bibinfo {note} {Starting from a maximally mixed density
  matrix, the quasi entropy of the time evolved density matrix at time $T$
  corresponds to imposing the twist boundary condition at $t=T$ boundary, and
  is given by $S^{(2)} = N \sigma L - \log \frac{T}{\sqrt{L}}$, $ T \gg L$ for
  periodic boundary condition for the chain. As expected, there are two phases
  corresponding to the volume-law and the area-law phases: when $\mu<J$, it
  takes an exponentially long time $\sqrt{L} e^{N \sigma L}$ to purify the
  state; when $\mu >J $, it takes a constant time $\sqrt{\xi}$ independent of
  system sizes to purify the state, where $\xi$ is the correlation length in
  the symmetric phase. The critical exponent is the same as the entanglement
  transition, as they differ only in boundary conditions.}\BibitemShut {Stop}%
\end{thebibliography}

\setcounter{secnumdepth}{3}
\setcounter{equation}{0}
\setcounter{figure}{0}
\renewcommand{\theequation}{S\arabic{equation}}
\renewcommand{\thefigure}{S\arabic{figure}}
\renewcommand\figurename{Supplementary Figure}
\renewcommand\tablename{Supplementary Table}
\newcommand\Scite[1]{[S\citealp{#1}]}
\makeatletter \renewcommand\@biblabel[1]{[S#1]} \makeatother

\newpage
\begin{widetext}

\section{Derivation of the effective action and the saddle-point equation of the monitored system}

As stated in the main text, the measurement on the four contours can be cast into (note that the measurement operator is Hermitian)
\bea
&& \sum_\nu w_\nu (K_\nu^{x,i})^{\otimes 2} \otimes (K_\nu^{x,i\dag})^{\otimes 2}\\
&=& (1- p) I^{\otimes4} + p [ (M_1^{x,i})^{\otimes 4} + (M_2^{x,i})^{\otimes 4} ] \\
&\approx& (1- p) +p \left(1 - \frac{s^2}2 \sum_{\alpha=1}^4 \pi^{+,\alpha}_{x,i} + s^4 \otimes_{\alpha=1}^4\pi^{+,\alpha}_{x,i} \right) \\
&\approx& \exp p  \left( - \frac{s^2}2 \sum_{\alpha} \pi^{+,\alpha}_{x,i}\right) \\
& = & \exp  \frac{\delta t \mu}{2}  \sum_\alpha i \psi_{x,L,i}^\alpha \psi_{x,R,i}^\alpha.
\eea
where we have used the relation $\pi^+_{x,a,j} + \pi^-_{x,a,j} = 1$ and also introduced $\alpha = 1,...,4$ to denote the four copies of the tensor product. To derive the above equation, we assume $ s \ll 1$ and keep orders up to $O(s^2)$. In the last line we introduce $ \mu = p s^2/\delta t $ and when the above limit is taken, $\mu$ is kept fixed. All the constants are neglected because they will not affect the dynamics. We arrive at (6) in the main text.

The derivation of $G$-$\Sigma$ action was firstly derived in~\cite{maldacena2016remarks}, and we outline the major steps in the derivation and refer the details to that paper. Let us look at a single chain first. The action on the four contours reads
\bea
    -I = \int dt \sum_\alpha \left( \sum_{x,i} \psi_{x,i}^\alpha (-1)^{\alpha+1} \partial_t \psi_{x,i}^\alpha + (-1)^\alpha i H[\psi^\alpha] \right),
\eea
where $\alpha=1,...,4$ denote the four contours, and $H[\psi] = H$ given in (1). Performing an average over the Gaussian variable using (2), it becomes
\bea
    -I &=& \sum_{\alpha, \beta,x} \int dt_1 dt_2 \Big( \sum_{i}  \frac12 \psi_i^\alpha (t_1) (-1)^{\alpha} \delta^{\alpha\beta}\delta(t_1-t_2)\partial_{t_2} \psi_i^\beta(t_2) \\
    && + (-1)^{\alpha+\beta+1} i^2 \frac{J \delta(t_1-t_2)}{4N} \sum_{i,j} \psi_{x,i}^\alpha(t_1) \psi_{x+1,j}^\alpha(t_1) \psi_{x,i}^\beta(t_2) \psi_{x+1,j}^\beta(t_2) \\
    && + (-1)^{\alpha+\beta+1} i^{q} \frac{U \delta(t_1-t_2)}{8q N^{q-1}} \sum_{j_1,...,j_q} \psi_{x,j_1}^\alpha(t_1)... \psi_{x,j_q}^\alpha(t_1) \psi_{x,j_1}^\beta(t_2)... \psi_{x,j_q}^\beta (t_2) \Big),
\eea
Now we introduce the bilocal fields $G_x^{\alpha\beta}(t_1,t_2)$ and $\Sigma_x^{\alpha\beta}(t_1,t_2)$. 
The bilocal field $G_x^{\alpha\beta}(t_1,t_2)$ is the correlation function of Majorana fermions,
\bea
    G_x^{\alpha\beta}(t_1, t_2) = \frac1N \sum_i \psi_{x,i}^\alpha(t_1) \psi_{x,i}^\beta(t_2),
\eea
and $\Sigma_x^{\alpha\beta}(t_1,t_2)$ is the self-energy of Majorana fermions, and is introduced through the following identity,
\bea
    1 = \int d\Sigma \exp \int dt_1 dt_2 \Big[ - \frac{N}2 \Sigma^{\alpha\beta}_x(t_1,t_2) \Big( G_x^{\alpha\beta}(t_1,t_2) - \frac1N  \sum_i \psi_{x,i}^\alpha(t_1) \psi_{x,i}^\beta(t_2)\Big) \Big].
\eea
By multiplying this identity and using $G_x^{\alpha\beta}$ to rewrite the coupling, we have
\bea
    -I &=& \sum_{\alpha, \beta,x} \int dt_1 dt_2 \Big( -\sum_{i} \frac12 \psi_i^\alpha \big( (t_1) (-1)^{\alpha+1} \delta^{\alpha\beta}\delta(t_1-t_2)\partial_{t_2} - \Sigma_{x}^{\alpha\beta}(t_1, t_2) \big) \psi_i^\beta(t_2) - \frac{N}2 \Sigma^{\alpha\beta}_x(t_1,t_2) G_x^{\alpha\beta}(t_1,t_2) \\
    && + N (-1)^{\alpha+\beta+1} \frac{J \delta(t_1-t_2)}{4} G^{\alpha\beta}_x(t_1,t_2) G^{\alpha\beta}_{x+1}(t_1,t_2) + N (-1)^{\alpha+\beta+1} \frac{U \delta(t_1-t_2)}{8q } [G^{\alpha\beta}_x(t_1,t_2)]^q \Big).
\eea
Now the action is quadratic in the Majorana field, so we can integrate over the Majorana fermions and get the $G$-$\Sigma$ action,
\bea
    - \frac{I}N = \frac12 \Tr \log \left( (-1)^{\alpha+1}  \partial_t - \Sigma_{x} \right)- \frac12 \int  \Sigma_{x}^{\alpha\beta} G_{x}^{\alpha\beta} 
     + \int \delta(t-t') \Big[ - \frac{(-1)^{\alpha + \beta}}{4}  \Big( J G_{x}^{\alpha\beta} G_{x+1}^{\alpha\beta} + \frac{U}{2q} (2G_{x}^{\alpha\beta})^{q} \Big) \Big]. 
\eea
A generalization to left and right chains, i.e., $G_{x,ab}^{\alpha\beta}(t_1,t_2) = \frac1N \sum_i \psi_{x,a,i}^\alpha(t_1) \psi_{x,b,i}^{\beta,i}(t_2) $ is straightforward.
Then combining with the measurement part, we arrive at the effective action in~(8). The saddle-point equation followed from (8) reads
\bea
    [G_x^{-1}]^{\alpha\beta}_{ab} &=& (-1)^{\alpha+1} \delta^{\alpha\beta} \delta_{ab}\partial_t - \Sigma_{ab,x}^{\alpha\beta}, \\
    \Sigma_{ab,x}^{\alpha\beta} &=& \delta(t-t') \Big[ \frac{- (-1)^{\alpha+\beta} \delta_{ab}}{2} \Big( J (G_{ab,x-1}^{\alpha\beta} + G_{ab,x+1}^{\alpha\beta}) + U(2 G_{ab,x}^{\alpha\beta})^{q-1} \Big) + i \mu \delta^{\alpha\beta} \frac{\delta_{aL}\delta_{bR} -\delta_{aR}\delta_{bL}}{2}\Big].
\eea

We consider the homogeneous solution in real space, i.e., $G_{ab,x}^{\alpha\beta} = \bar G_{ab}^{\alpha\beta}$ and $\Sigma_{ab,x}^{\alpha\beta} = \bar \Sigma_{ab}^{\alpha\beta}$. To get the solution, we focus on two contours, $\alpha, \beta = 1,2$, because the boundary condition in $\Tr(\rho)^2$ is to connect 1 to 2 and connect 3 to 4 separately. 
If the evolution is unitary, the correlation between two contours will be $G^{12}_{aa}(t,t)=-\frac12$ because the forward and backward evolution cancels. The effect of non-Hermitian couplings is to decrease this correlation, and therefore, we assume the correlation is given by $G^{12}_{aa}(t,t)=- \frac{\zeta}2$. 
(S13) shows that the self-energy is a function of time difference and proportional to Dirac delta function. It is convenient to work in frequency space $\bar \Sigma_{ab}^{\alpha\beta}(t_1,t_2) = \int \frac{d\omega}{2\pi} \bar \Sigma_{ab}^{\alpha\beta}(\omega) e^{-i\omega (t_1-t_2)} $. According to (S13), we have $\bar \Sigma(\omega) = - \frac12 (J \zeta + U \zeta^{q-1})  i \sigma^y  - \frac{\mu}2 \tau^y$, where $\sigma$ ($\tau$) acts on the $1,2$ contours (the $L,R$ chains). 
Using (S12), the Green's function at the equal time reads
\bea
    \bar G(t,t) = \frac{1}{2} \left( - \frac{J\zeta^2+U\zeta^q}{\sqrt{(J \zeta^2 + U \zeta^{q} )^2+ \mu^2 \zeta^2}} i \sigma^y + \frac{\mu \zeta}{\sqrt{(J \zeta^2 + U \zeta^{q} )^2+ \mu^2 \zeta^2}} \tau^y  \right).
\eea 
Requiring $\bar G^{12}_{aa}(t,t) = - \frac\zeta2$, we have 
\bea
        (1-\zeta^2)(J + U \zeta^{q-2})^2 = \mu^2,
\eea
which leads to (15) in the main text. 
For noninteracting case, $U=0$, the solution is $\zeta= \sqrt{J^2 - \mu^2}$, which is (9) in the main text.

\section{Derivation of Goldstone mode effective action}

We consider the fluctuation away from the saddle point solution (9) at $\mu < J$. 
First notice that $G_{LR,x}^{\alpha\alpha}$ is a linear term in the action, so it can be integrated out to enforce $\Sigma_{LR,x}^{\alpha\alpha} =  \frac{i \mu}2$. 
Then we consider the fluctuations $ G_{aa,x}^{\alpha\beta}(t_1,t_2) = \bar G_{aa}^{\alpha\beta}(t_1,t_2) + \delta G_{aa,x}^{\alpha\beta}(t_1,t_2)$ and $\Sigma_{aa,x}^{\alpha\beta}(t_1,t_2) = \bar\Sigma_{aa}^{\alpha\beta}(t_1,t_2) + \delta \Sigma_{aa,x}^{\alpha\beta}(t_1) \delta(t_1-t_2)$, $a=L, R$. 
Expanding the tr log term in (8) leads to the kernel of $\delta \Sigma$, i.e., 
\bea \label{eq:kernel}
    \frac12 \Tr \log \left( (-1)^{\alpha+1}  \partial_t - \Sigma_{x} \right) \approx - \frac14 \int \frac{d\omega d\Omega}{(2\pi)^2} \sum_{a,\alpha\beta\gamma\delta}\delta \Sigma^{\alpha\beta}_{aa,x}(\Omega) \bar G_{aa}^{\beta\gamma}(\omega ) \bar G_{aa}^{\delta\alpha}(\omega+ \Omega) \delta\Sigma^{\gamma\delta}_{aa,x}(-\Omega),
\eea
where $\delta \Sigma_{aa,x}^{\alpha\beta}(t) = \int \frac{d\omega}{2\pi} \delta \Sigma_{aa,x}^{\alpha\beta}(\omega) e^{-i\omega t}$. 
The kernel can be brought into decoupled sectors by a basis transformation. The calculation is straightforward but tedious, so we only present the main result.
The interesting part is the following,
\bea
\frac12 \int_\Omega \sum_x \hat \sigma_x(\Omega) \left( \ba{cccc} - \frac{J^2 - \mu^2}{J(J^2 + \Omega^2)} &  -i \sqrt{1 - (\frac{\mu}J)^2}\frac{\Omega}{ J^2 + \Omega^2} \\
-i \sqrt{1 - (\frac{\mu}J)^2}\frac{\Omega}{ J^2 + \Omega^2} & \frac{J}{J^2 + \Omega^2} \ea \right) \hat \sigma_x(-\Omega),
\eea
where $\int_\Omega \equiv \int \frac{d\Omega}{2\pi}$, $\hat \sigma_x \equiv \sum_a \left(  \frac12 (\delta \Sigma^{13}_{aa,x} + \delta \Sigma^{24}_{aa,x} ) ,  \frac12 (\delta \Sigma^{14}_{aa,x} + \delta \Sigma^{23}_{aa,x} ) \right)$ decouples from the other fluctuations. 
The reason to consider the above component is that $ \sigma_{2,x}$ corresponds to the Goldstone mode and $ \sigma_{1,x}$ cannot be neglected because it couples to the Goldstone mode.

Next we expand the other terms in (8). The second term in (8), i.e., 
\bea 
    - \sum_{a,x} \frac12 \int dt dt'  \Sigma_{aa,x}^{\alpha\beta}(t,t') G_{aa,x}^{\alpha\beta} (t,t'), 
\eea
leads to the coupling between $\hat \sigma_x$ and $\delta G_{aa,x}$ and the term
\bea
 	&  \sum_{ab,x} \int dt dt' \delta(t-t')  [- \frac{(-1)^{\alpha + \beta}}{4} \delta_{ab} J G_{ab,x}^{\alpha\beta}(t,t') G_{ab,x+1}^{\alpha\beta}(t,t')],
\eea
contributes to the kinetic term of $\delta G_{aa,x}$. Here we are interested in noninteracting case so $U=0$. Integrating over $\hat \sigma_x$, we arrive at
\bea \label{eq:SYK2_effective_action1}
\frac{-I_\text{eff}}{N} = \frac12 \sum_k \int_\Omega \hat \varphi_k(\Omega) \left( \ba{cccc} \frac{J^3}{J^2 - \mu^2} - J_k & \frac{i J \Omega}{\sqrt{J^2 - \mu^2}} \\
-\frac{i J \Omega}{\sqrt{J^2 - \mu^2}} & -J+ J_k \ea \right) \hat \varphi_{-k}(-\Omega),
\eea
where $\hat \varphi_x \equiv \sum_a \left(  \frac12 (\delta G^{13}_{aa,x} + \delta G^{24}_{aa,x} ),\frac12 (\delta G^{14}_{aa,x} + \delta G^{23}_{aa,x}) \right)$, or more explicitly in components, $ \varphi_{1,x}= \sum_a  \frac12 (\delta G^{13}_{aa,x} + \delta G^{24}_{aa,x} ), \varphi_{2,x}=\sum_a \frac12 (\delta G^{14}_{aa,x} + \delta G^{23}_{aa,x}) $.  $J_k \equiv J \cos k$, and the convention for Fourier transformation is $ \hat \varphi_x = \frac1{\sqrt{L}} \sum_k \hat \varphi_k e^{i k x}$.
We further integrate out the $\varphi_{1,k}$ field as it is gaped to get the effective action
\bea \label{eq:SYK2_effective_action2}
\frac{-I_\text{eff}}{N} = \frac12 \sum_k \int_\Omega \varphi_{2,k}(\Omega) \left( -\frac{J }{\mu^2}\Omega^2 + J_k - J \right) \varphi_{2,-k}(-\Omega).
\eea
It seems the action is intact at $\mu = J$ contradicting our proposal that transition occurs at $\mu = J $.
However we should note that $\varphi_{2,k}$ is related to the Goldstone mode nontrivially. 
Namely~(13) indicates the $\varphi_{2,k}$ is related to the Goldstone mode $\theta(t)$ nontrivially by
\bea 
    \varphi_{2,x}(t) =  \sqrt{1- \left(\frac{\mu}J \right)^2} \theta_x(t).
\eea
Using this relation, we finally have the effective theory for the Goldstone mode $\theta$ in (14),
\bea
\frac{-I_\text{eff}}{N} = \frac12  \left( 1- \left(\frac{\mu}J \right)^2 \right) \sum_k \int_\Omega \theta_k(\Omega) \left( \frac{J}{\mu^2} \Omega^2 + J_k - J \right) \theta_{-k}(-\Omega).
\eea

It is also interesting to look at $\varphi_{1,x} = \sum_a \frac12 (\delta G^{13}_{aa,x} + \delta G^{24}_{aa,x})$. Due to the presence of the gapless Goldstone mode, the correlation function of $\varphi_{1,x}$ becomes algebraic, i.e.,
\bea
    \langle \varphi_{1,k}(\Omega) \varphi_{1,q}(-\Omega)  \rangle \approx \frac{(J^2-\mu^2) k^2 \delta_{q,-k}}{2J (\Omega^2 + \frac{\mu^2}2 k^2)}, \quad \langle \varphi_{1,r}(t) \varphi_{1,0}(t)  \rangle \propto \frac1{r^2},
\eea
where the first equation is obtained from (\ref{eq:SYK2_effective_action1}) with an expansion at $k \approx 0$. The second equation is the equal time correlation function in the real space. It is obtained from the first equation by the Fourier transform $\varphi_{1,r}(t) \equiv \int \frac{dk d\Omega}{(2\pi)^2} \varphi_{1,k}(\Omega) e^{ikr-i \Omega t}$, where a continuum limit of lattice is made. The power-law correlations can be observed in equal-time squared correlation function of fermions~\cite{part_one, chen2020emergent, alberton2021entanglement}.

\section{Derivation of $Z_4$ effective action}

We consider the fluctuation away from the symmetric saddle point solution~(9) at $\mu \ge J$. 
The derivation is similar to the effective Goldstone action shown in the previous section. Again we consider fluctuations, $ G_{x,aa}^{\alpha\beta}(t_1,t_2) = \bar G_{aa}^{\alpha\beta}(t_1,t_2) + \delta G_{x,aa}^{\alpha\beta}(t_1,t_2)$ and $  \Sigma_{x,aa}^{\alpha\beta}(t_1,t_2) = \bar \Sigma_{aa}^{\alpha\beta} + \delta \Sigma_{x,aa}^{\alpha\beta}$ and $\delta G_{aa}^{\alpha\beta}$, $a=L, R$.
Expanding $\delta \Sigma$  in the trace log term in (8) and the result is (\ref{eq:kernel}).
It can again be brought into decoupled sectors. Because they serve as an order parameter we focus on the components $(\delta \Sigma^{12}_{aa},\delta \Sigma^{34}_{aa},\delta \Sigma^{14}_{aa},\delta \Sigma^{23}_{aa})$ whose kernel is given by
\bea \label{eq:kernel}
K = \frac{\mu}{4(\mu^2 + \Omega^2)} 1_{4\times 4} \otimes \left( \ba{cccc} 1 & 1 \\ 1 & 1 \ea \right),
\eea
where the first matrix is in the basis of these four components and the second is in the basis of the $L$ and $R$ chains.
It is apparent that there are four zero modes and integrating them out will lead to the following constraints,
\bea
\delta G^{12}_{RR} = \delta G^{12}_{LL},\quad \delta G^{34}_{RR} = \delta G^{34}_{LL}, \quad \delta G^{14}_{RR} = \delta G^{14}_{LL}, \quad \delta G^{23}_{RR} = \delta G^{23}_{LL}.
\eea
Thus there are four independent fields left $(\delta G^{12},\delta G^{34},\delta G^{14},\delta G^{23})$ where we suppress the subscript.

Now it is a straightforward task to integrate out the rest fluctuations with nonzero kernel in~(\ref{eq:kernel}). 
The nontrivial sector that are related to the $C_4$ symmetry is span by $(\phi_1, \phi_2) = (\delta G^{12} + \delta G^{34},\delta G^{14}+ \delta G^{23})$ which transforms as a vector under the $C_4$ operator.
The effective theory reads
\bea \label{eq:Z4_model1}
& \frac{I_\text{eff}}{N} = \sum_{i=1,2} \sum_k \int_{\Omega} \phi_{i,k}(\Omega) \left( \frac{\Omega^2}{2\mu} + \frac{J}4 k^2 + \frac{\mu-J}2 \right) \phi_{i,-k}(-\Omega)  +  \sum_x \int_{t} \left( - \frac{2^q U}{16 q} ( \phi_{1,x}^q + \phi_{2,x}^q )+ V (\phi_{1,x}^2 + \phi_{2,x}^2 )^{q/2} \right),
\eea
where $\int_t \equiv \int dt$ and we include the last term which should be obtained by expanding the trace log term to higher orders for stability of the theory. 
We will focus on $q=4$ in the following. 
In this case $V \approx \frac{\mu}8$ near the transition point inferred from the condition (16) of the parameter $\zeta$ for continuous transitions at $\mu = J$ and the fact that the saddle point solution depends only on $\mu$ for $\mu \ge J$.
This leads to (17) in the main text.

We have expanded the action around the symmetric saddle point solution, but we may also wish to explore the ordered phase by expanding around the asymmetric saddle point solution. 
Similar to we have done to get the noninteracting case~(\ref{eq:SYK2_effective_action1}), we have the following effective action
\bea
\frac{-I_\text{eff}}{N} = \frac12 \sum_k \int_\Omega \hat \varphi_k(\Omega) \left( \ba{cccc} \frac{(y^2 J^2 + \mu^2)^{3/2}}{2J^2 y^2} - J_k & \frac{i \sqrt{y^2 J^2 + \mu^2} \Omega}{yJ} \\
-\frac{i \sqrt{y^2 J^2 + \mu^2} \Omega}{yJ} & -\sqrt{y^2 J^2 + \mu^2} + J_k \ea \right) \hat \varphi_{-k}(-\Omega),
\eea
where the parameter $y$ is given by
\bea
y= \frac{y}{\sqrt{y^2 + \tilde \mu^2}} + \tilde U \left(\frac{y}{ \sqrt{y^2 + \tilde \mu^2}} \right)^{q-1}.
\eea 
For small interaction strength, $y$ is given by
\bea
y = \sqrt{1- \tilde \mu^2} + \tilde U (1- \tilde \mu^2)^{\frac{q-3}2} + O(\tilde U^2).
\eea
Now we expect the Goldstone mode acquires a finite mass proportional to $ U$ due to the lowering of symmetry from $O(2)$ to $C_4$. 
Indeed by integrating out the $\varphi_1$ we have
\bea
\frac{-I_\text{eff}}{N} = \frac12 \sum_k \int_\Omega \varphi_{2,k}(\Omega) \left( - \sqrt{y^2 J^2 + \mu^2 } + J_k - \frac{\Omega^2}{\sqrt{y^2 J^2 + \mu^2 } - \frac{y^2 J^2}{y^2 J^2 + \mu^2 } J_k} \right) \varphi_{2,-k}(-\Omega).
\eea
The mass $ J - \sqrt{y^2 J^2 + \mu^2 } $ vanishes when $U = 0$, and we restore~(\ref{eq:SYK2_effective_action2}).
For small interactions, the mass can be simplified as $J - \sqrt{y^2 J^2 + \mu^2 } \approx U(1-\tilde \mu^2)^{\frac{q}2-1}$.
It is apparent that the interaction reduces the symmetry and renders the Goldstone mode gaped.

\section{Numerical calculation of R\'enyi entropy and mutual information}

In this section, we give more details about how to numerically calculate the R\'enyi entropy in Fig.3 and 4. 
The calculations have been discussed in many previous works~\cite{penington2019replica, chen2020replica, zhang2020entanglement, liu2021non, jian2020note}.
We discuss the main changes in this paper, and refer the details of the calculation to these works. 
The starting point is the following expression of R\'enyi-2 entropy,
\bea
    \exp(-S_A^{(2)}) =  \frac{\mathbb E \Tr(\rho_{A}^2)}{\mathbb E\Tr(\rho)^2},
\eea
where $\rho$ is unnormalized density matrix ($\rho_A$ is unnormalized reduced density matrix of subsystem $A$), and $\mathbb E$ denotes average of Brownian variables and the quantum trajectories. 
Both the numerator and denominator can be written as a path integral, where the effective action is,
\bea \label{eq:action}
    && - \frac{I[G_x,\Sigma_x; F_x]}N = \frac12 \Tr \log \left( \delta_{ab} F_x(s,s') - \Sigma_{ab,x}(s,s') \right)- \frac12 \int ds ds' \Sigma_{ab,x}(s,s') G_{ab,x} (s,s')  \nn \\
    && + \int ds ds' g(s,s') \Big[  \frac{f(s)f(s')}{4} \delta_{ab} \Big( J G_{ab,x}(s,s') G_{ab,x+1}(s,s')  + \frac{U}{2q} (2G_{ab,x}(s,s') )^{q} \Big) + \frac{i \mu}2 \delta(s-s') G_{LR}(s,s')  \Big],
\eea
where $0<s,s'<4T$ is a label of four contours: the forward contours are $s\in (0,T) \cup (2T, 3T)$ and the backward contours are $s\in (T,2T) \cup (3T, 4T)$. The bilocal fields $G$ and $\Sigma$ do not have the contour index $\alpha$ or $\beta$ since $s$ and $s'$ label the contours.
The function $f(s)$ characterises the arrow of times in the unitary evolution,
\bea 
f(s) = \begin{cases} i, & s\in (0,T) \cup (2T, 3T) \\
-i, & s\in (T,2T) \cup (3T, 4T) \end{cases}.
\eea
$g(s,s') = \sum_{\alpha=0}^1 \delta(|s-s'|-2\alpha T)+ \sum_{\alpha=1}^3 \delta(s+s'-2\alpha T)$ is the Brownian correlation on the four contours. $F_x(s,s')$ is a function that will be specified later.

Because of the large-$N$ structure, we can solve the Dyson-Schwinger equation numerically and calculate the onshell action. The Dyson-Schwinger equation followed from the action reads 
\bea \label{eq:dyson}
 [G_x^{-1}]_{ab}(s,s') &=& \delta_{ab} F_x(s,s') - \Sigma_{ab,x}^{\alpha\beta}(s,s'), \\
\Sigma_{ab,x}(s,s') &=& g(s,s') \Big[ \frac{f(s)f(s') \delta_{ab}}{2} \Big( J (G_{ab,x-1}(s,s')) + G_{ab,x+1}(s,s')) + U(2 G_{ab,x}(s,s'))^{q-1} \Big) \nn \\
&& + i \mu \delta(s-s') \frac{\delta_{aL}\delta_{bR} -\delta_{aR}\delta_{bL}}{2}\Big].
\eea
Then the R\'enyi entropy is given by
\bea \label{eq:onshell}
    \exp(-S_A^{(2)}) =  \frac{\mathbb E \Tr(\rho_{A}^2)}{\mathbb E\Tr(\rho)^2} \approx \frac{e^{-I_{\text{onshell}}(A)}}{e^{-I_{\text{onshell}}(\emptyset)}}, 
\eea
where $I_{\text{onshell}}(A)$ and $I_{\text{onshell}}(\emptyset)$ are defined as
\bea
    I_{\text{onshell}}(A) = I[\bar G_x, \bar \Sigma_x; F_{A,x}], \quad I_{\text{onshell}}(\emptyset) = I[\bar G_x, \bar \Sigma_x; F_{\emptyset,x}].
\eea
Here $\bar G_x$, $\bar \Sigma_x$ denotes the numerical solutions from iteration, and $F_{A,x}$ and $F_{\emptyset, x}$ are defined in the following to account for the boundary condition.

The difference between the numerator and denominator in~(\ref{eq:onshell}) is exactly the boundary condition. 
To be concrete, we double the full system and choose the initial state to be thermofield double state~\cite{penington2019replica, jian2020note}.
Because of the Brownian nature of the model, we consider infinite temperature thermofild double state, the details of which can be found in~\cite{jian2020note}. 
As we are interested in R\'enyi entropy of subsystem $A$, the twist boundary condition is applied in subsystem $A$. 
To incorporate the different boundary conditions, the $F$ is chosen differently.
We define
\bea
    (F^{(0)})^{-1}(s,s') = \frac12 \sgn(s-s'), \quad s,s' \in (0,2T) \text{ or } s,s' \in (2T,4T), \\
    (F^{(1)})^{-1}(s,s') = \frac12 \sgn(s-s'), \quad s,s' \in (T,3T) \text{ or } s,s' \in (0,T)\cup(3T,4T).
\eea
In the denominator, there is no twist boundary condition, 
while in the numerator, the twist boundary condition is implemented in subsystem $A$. To account for these boundary conditions, the function $F$ is given by
\bea
    F_{\emptyset, x}(s,s') &=& F^{(0)}(s,s'), \quad \forall x, \\
    F_{A,x}(s,s') &=& \begin{cases} F^{(0)}(s,s'), & x \notin A \\ 
    F^{(1)}(s,s'), & x \in A \end{cases}.
\eea

Having specified all the functions in~(\ref{eq:dyson}), the Dyson-Schwinger equation can be solved numerically by discretizing $s,s'$, and iterating the equation, the details of which can be found in~\cite{maldacena2016remarks, penington2019replica, jian2020note}. The discretization implemented in our calculation is $400$ to $800$, which is converged. Other parameters are specified in the caption of each figure.
To calculate the mutual information, we use the definition $I_{AB}=S^{(2)}_{AB}-S^{(2)}_A-S^{(2)}_B$, and calculate the R\'enyi entropy on the right-hand side. The quantity $S^{(2)}_{AB}$ is given as
\bea
    \exp(-S^{(2)}_{AB}) \approx \frac{e^{-I_{\text{onshell}}(A\cup B)}}{e^{-I_{\text{onshell}}(\emptyset)}},
\eea
where $I_{\text{onshell}}(A\cup B) = I(\bar G_x, \bar \Sigma_x; F_{A\cup B, x})$, 
\bea
    F_{A\cup B,x}(s,s') &=& \begin{cases} F^{(0)}(s,s'), & x \notin A\cup B \\ 
    F^{(1)}(s,s'), & x \in A\cup B \end{cases}.
\eea

\end{widetext}

\end{document}